\documentclass[hyper,notoc,a4paper]{JHEP3}
\usepackage[intlimits]{amsmath}
\usepackage[english]{babel}
\usepackage{setspace}
\title{Matrix and vector models \\ in the strong coupling limit}
\author{D.V.Bykov, A.A.Slavnov\\ M.V.Lomonosov Moscow State University \newline Faculty of Physics, Leninskie gory, bld. 1-2, GSP-2 119992 Moscow, Russia\\
V.A.Steklov Mathematical Institute,
\newline
Gubkina str., bld. 8, GSP-1 119991 Moscow, Russia \\
\email{dbykov@mi.ras.ru;} \email{slavnov@mi.ras.ru}}
\date{1.07.2007}
\abstract{In this paper we consider matrix and vector models in the
large $N$ limit ($N\times N$ matrices and vectors with $N^{2}$
components). For the case of zero-dimensional model (D=0) it is
proved that in the strong coupling limit $g\to \infty$ statistical
sums of both models coincide up to a coefficient. This is also true
for $D=1$.}
\newcommand{\tr}{{\rm tr}}

\newpage
\def\XXint#1#2#3{{\setbox0=\hbox{$#1{#2#3}{\int}$}
     \vcenter{\hbox{$#2#3$}}\kern-0.65\wd0}}

\begin{document}
\large \setstretch{1}
\section{Introduction}
Matrix models have applications in various branches of physics, but
for us the main motivation will be the possibility of using them in
quantum field theory. In particular, as it has been shown by
\mbox{'t Hooft}~\cite{hooft}, with a proper normalization of
coupling constants the expansion of Green's functions (and the
partition function) in $\frac{1}{N}$ corresponds in terms of ribbon
Feynman diagrams to the expansion in the genuses of Riemannian
surfaces, on which this graph can be drawn without
self-intersections. However, even if the genus of the surface is
fixed, there are infinitely many corresponding ribbon graphs and in
the majority of interesting cases, unfortunately, there's still no
algorithm for their summation.

The limit $N \to \infty$ is also interesting from the point of view
of the Ads/CFT correspondence ~\cite{ads}. In this case one should
consider the large $g$ limit of the sum of planar diagrams
($N\to\infty$). This problem is far from being solved not only for
gauge theories but also for the simpler case of scalar matrix
models. Exceptions are cases $D=0$ and $D=1$, where the algorithm of
planar diagram summation has been found \cite{brezin} for real and
hermitian matrix models. Later (at $D=0$) it was generalized for
complex matrix models \cite{makeenko}. A review of the
state-of-the-art condition of the theory of zero-dimensional matrix
models can be found in \cite{chekhov}.

Vector and matrix models at $D=0,1$ reveal remarkable similarity in
the limit $g\to\infty$. The dependence of the partition function on
the coupling constant in this limit is the same for vector and
matrix models. This suggests that such coincidence can also take
place in higher dimensions.

However, in this paper we will restrict ourselves to the
consideration of a $D=0$ complex matrix model, i.e. we consider the
following partition function:
$$
\mathcal{Z}(g)=\int\,d\varphi^{\dag}\,d\varphi\,\exp[-\frac{1}{2}
{\rm tr}(\varphi^{\dag}\varphi)-\frac{g}{4N}{\rm
tr}((\varphi^{\dag}\varphi)^{2})],
$$
where $\varphi$ is a $N\times N$ matrix. The notation
$d\varphi^{\dag}\,d\varphi$ should be interpreted in the following
way:
$$d\varphi^{\dag}\,d\varphi \equiv D\varphi \equiv \prod\limits_{i,j}\,d{\rm
Re}\varphi_{ij}\,d{\rm Im}\varphi_{ij}.$$ Free energy is defined by
$$
\mathcal{E}=- \frac{\ln \mathcal{Z}_{BV}}{N^{2}}.
$$
We will show that at $D=0$ in the strong coupling limit ($g\to
\infty$) the leading asymptotics of the free energy of this matrix
model coincides with the leading asymptotics of the free energy of
the corresponding vector model. At $D=1$ this is also true, but a
more thorough consideration of this case is necessary. Of course,
especially interesting are the cases $D \geq 2$, but there the
question of whether the described hypothesis is true is still open.
We would like to emphasize from the beginning that at $D \geq 2$
ultraviolet divergences appear both in matrix and vector models,
therefore it only makes sense to speak about properly regularized
theories.

Notice that in all dimensions linear vector models with \sloppy
interaction $N^{2}F(\frac{1}{N^{2}}\tr(\varphi^{\dag}\varphi))$ are
exactly soluble, i.e. Green's functions of singlet variables can be
found in explicit form, the analogy between matrix and vector models
which supposedly takes place can help in the analysis of (much more
complicated) matrix models.
\section{D=0 bivector models}
For the comparison of matrix and vector models we first calculate
the partition function of the vector model. Let's explain what we
mean by vector and bivector models. The distribution $\rho$ of
random variables $x_{a}$ ($a=1...M$) in a vector model, by
definition, depends only on the square of the vector $\mathbf{x}$:
$\rho=\rho(x_{a}^{2})$.

Consider the partition function
\begin{equation}\label{21}
\mathcal{Z}_{BV} = \int\,D\varphi \,\exp{\left[-\frac{1}{2}\tr
(\varphi^{\dag}\varphi)+N^{2} F(\frac{1}{N^{2}}\tr
(\varphi^{\dag}\varphi))\right]}.
\end{equation}
Although formally this model is a matrix model, in fact it is a
vector model, because if we introduce a vector $y_{a}$ with $M=2
N^{2}$ components $$(y_{i}={\rm Re}\varphi_{i,1}, y_{N^{2}+i}= {\rm
Im}\varphi_{i,1}, i \leq N),\; (y_{i}={\rm Re}\varphi_{i-N,2},
y_{N^{2}+i}={\rm Im}\varphi_{i-N,2},N < i \leq 2N) ...,$$ then the
distribution turns out to be just a function of $y_{a}^{2}$. We call
such models bivector models. They are characterized by the fact that
their symmetry group is much wider than in the corresponding matrix
model: for example, in this case the matrix model has a symmetry
group $U(1)\times SU(N)_{L}\times SU(N)_R$ (multiplication by a
phase factor, left and right matrix multiplication
\footnote{Multiplication by a phase factor does not reduce to a
combination of left and right multiplication by unimodular matrices
$\varphi'=U_{1}\varphi U_{2}^{\dag}$. Indeed, suppose it is possible
and set $\varphi=1$. Then $U_{1}U_{2}^{\dag}=e^{i\alpha}I$, which
obviously is not true if $\alpha \neq 0$.}), whereas the vector
model is invariant under $U(2 N^{2})$. This large symmetry is what
lets us find an exact solution for (bi)vector models.

First we want to find the $N\to \infty$ limit of the free energy.
Inserting a unity into the integrand of (\ref{21}) we obtain:
\begin{equation}\label{22}
\mathcal{Z}_{BV} = \int\,D\varphi \,d\rho
\,\delta(\rho-\tr(\varphi^{\dag}\varphi))\,\exp{\left[-\frac{1}{2}\tr
(\varphi^{\dag}\varphi)+N^{2} F(\frac{1}{N^{2}}\rho)\right]}.
\end{equation}
Let us write the Fourier transformation for the delta-function:
\begin{equation}\label{23}
\mathcal{Z}_{BV} = (2\pi)^{-1}\int\,D\varphi \,d\rho\,d\lambda\,
\exp{\left[-\frac{1}{2}\tr (\varphi^{\dag}\varphi)+N^{2}
F(\frac{1}{N^{2}}\rho)+i\lambda(\rho-\tr(\varphi^{\dag}\varphi))\right]}.
\end{equation}
The integral over $\varphi$ is gaussian, moreover it degenerates
into a product of independent integrals because the components of
the matrix $\varphi$ are not linked:
$\tr(\varphi^{\dag}\varphi)=\sum\limits_{i, j}|\varphi_{ij}|^{2}$.
As a result, the integral over $\varphi$ can be taken, and we get:
\begin{equation}\label{24}
\mathcal{Z}_{BV} = (2\pi)^{-1}\int \,d\rho\,d\lambda\,
\left(\frac{\pi}{1/2+i\lambda}\right)^{N^{2}}\, \exp{\left[N^{2}
F(\frac{1}{N^{2}}\rho)+i\lambda\rho\right]}.
\end{equation}
The integral over $\lambda$ can be taken in an explicit way, because
we can (under the condition $\rho>0$, which follows from the
definition $\rho=\tr(\varphi^{\dag}\varphi)$) close the contour in
the upper half-plane, and inside the contour there's a pole
$\lambda=\frac{i}{2}$ of order $N^{2}$. The integral equals $2\pi i
\times$(residue in this pole). Thus,
\begin{equation}\label{25}
\int
\,d\lambda\,\frac{e^{i\lambda\rho}}{(1+2i\lambda)^{N^{2}}}=\frac{1}{(N^{2}-1)!}\,\frac{2\pi}{\rho}\,\left(\frac{\rho}{2}\right)^{N^{2}}\,e^{-\frac{\rho}{2}}.
\end{equation}
We obtain
\begin{equation}\label{26}
\mathcal{Z}_{BV} =
\frac{\pi^{N^{2}}}{(N^{2}-1)!}\,\int\,d\rho\,\frac{1}{\rho}\rho^{N^{2}}e^{N^{2}F(\frac{\rho}{N^{2}})-\frac{\rho}{2}}.
\end{equation}
Next we make a change $\rho \rightarrow N^{2}\rho$ to reduce the
integral to the case, in which the steepest descent method is
applicable:
\begin{equation}\label{27}
\mathcal{Z}_{BV} = \frac{(N^{2}
\pi)^{N^{2}}}{(N^{2}-1)!}\,\int\,d\rho\,\frac{1}{\rho}\,e^{N^{2}(\ln\rho+F(\rho)-\frac{\rho}{2})}.
\end{equation}
The stationary point equation is
\begin{equation}\label{28}
\rho_{0}:\; \frac{1}{\rho_{0}}-\frac{1}{2}+F'(\rho_{0})=0.
\end{equation}
Now we specify the potential: $F(\rho)=-\frac{g}{4}\rho^{2}$. Then
the equation takes the form
\begin{equation}\label{29}
\frac{1}{\rho_{0}}-\frac{1}{2}-\frac{g}{2}\rho_{0}=0.
\end{equation}
A positive solution of the equation is
\begin{equation}\label{30}
\rho_{0}=\frac{1}{2g}(\sqrt{1+8g}-1)
\end{equation}
We get the following expression for the partition function in the
large $N$ limit:
\begin{equation}\label{31}
\mathcal{Z}_{BV} = \frac{(N^{2} \pi)^{N^{2}}}{(N^{2}-1)!}
\frac{1}{\rho_{0}}
\left(\frac{2\pi}{N^{2}g\left(\frac{1}{2}+\frac{1}{2-\rho_{0}}\right)}\right)^{1/2}e^{N^{2}(\ln\rho_{0}-\frac{\rho_{0}}{2}-g\frac{\rho_{0}^{2}}{4})}
\end{equation}
Using Stirling's formula for the factorial ($n!\approx
(\frac{n}{e})^{n} \sqrt{2\pi n}$),
\begin{equation}\label{32}
\mathcal{Z}_{BV} = (e\cdot \pi)^{N^{2}} \frac{1}{\rho_{0}}
\left(\frac{g}{2}+\frac{g}{2-\rho_{0}}\right)^{-1/2}e^{N^{2}(\ln\rho_{0}-\frac{\rho_{0}}{2}-g\frac{\rho_{0}^{2}}{4})}
\end{equation}
From (\ref{30}) it is clear that in the limit $g\rightarrow
\infty:\; \rho_{0}\sim(\frac{2}{g})^{1/2}\rightarrow 0$. Therefore
the only growing contribution to the free energy is given by the
logarithm, and
\begin{equation}\label{33}
\mathcal{E}_{0} \, \underset{g\to \infty}{\rightarrow} \,
\frac{1}{2}\ln(g).
\end{equation}
\section{Solution of the D=0 matrix model with ${\rm tr} ((\varphi^{\dag}\varphi)^{2}) $-interaction}
In this section we consider a model which has the $N\times N$
complex matrix $\varphi$ as its dynamical variable. We will consider
the partition function
\begin{equation}\label{3.1}
\mathcal{Z}=\int\,d\varphi^{\dag}\,d\varphi\,\exp[-\frac{1}{2} {\rm
tr}(\varphi^{\dag}\varphi)-\frac{g}{4N}{\rm
tr}((\varphi^{\dag}\varphi)^{2})].
\end{equation}
Correlation functions of operators
$\tr((\varphi^{\dag}\varphi)^{m})$ were obtained in \cite{makeenko}.
All planar Green's functions are constructed with the help of the
eigenvalue distribution function $u(\lambda)$, defined on a segment
$\lambda \in [0;\sqrt{z}]$ (here $\lambda$ should be interpreted as
the modulus of an eigenvalue, so $\lambda \geq 0$), which has the
following form:
\begin{equation}\label{3.2}
u(\lambda)=\frac{1}{2
\pi}\left(1+\frac{gz}{2}+g\lambda^{2}\right)\sqrt{z-\lambda^{2}},
\end{equation}
where $z=\frac{2}{3g}(\sqrt{1+24g}-1)$. Note that as $g\to \infty \;
z \to 0$. At the same time the area under the plot $u(\lambda)$ is
always equal to unity: $\int\,u(\lambda)\,d\lambda =1$. As a result,
obviously
\begin{equation}\label{3.3}
\underset{g\to\infty}{\lim}\,u(\lambda) = \delta(\lambda),
\end{equation}
where the limit should be understood, of course, in the sense of
distributions. One can see that in the limit $g \to \infty$ all
eigenvalues tend to approach zero. Nevertheless, it's not possible
to use directly the asymptotics (\ref{3.3}) in order to obtain the
partition function at large $g$, as in the expression for the free
energy there's a term $\int \,d\lambda\,u(\lambda)\,\ln{\lambda},$
and the logarithm is singular at $\lambda=0$. We propose another
method for the solution of a complex matrix model, which is free
from the mentioned drawback, and, using this method, we will prove
at $D=0$ the statement presented above about the equality of free
energies. First we make in the integral the Hubbard-Stratonovich
transformation of the second term in the exponent:
\begin{equation}\label{m1}
 \mathcal{Z} = \int\,d\varphi^{\dag}\,d\varphi\,d\eta\,\exp\left[-\frac{1}{2} {\rm
tr}(\varphi^{\dag}\varphi)-\tr \,\eta^{2}+i \sqrt{\frac{g}{N}}\tr
(\eta \varphi^{\dag}\varphi)\right],
\end{equation}
where $\eta$ is a hermitian matrix. Now we make the gaussian
integration \mbox{over $\varphi$:}
\begin{equation}\label{3.5}
 \mathcal{Z} = \int \,d\eta\, \exp{\left[-\tr \, \eta^{2} - N \tr \ln\left[\frac{1}{2}I-i\sqrt{\frac{g}{N}}\eta \right]
\right]}.
\end{equation}
After passing to the integration over (real) eigenvalues
$\lambda_{k}$ of a hermitian matrix $\eta$ the integral can be
calculated by the stationary phase method. Upon varying the exponent
over $\lambda_{k}$ we get the following equations:
\begin{equation}\label{m2}
  \sum\limits_{j=1}\,^{'} \, \frac{2}{\lambda_{k}-\lambda_{j}}= 2\lambda_{k}+\frac{i\sqrt{gN}}{-1/2+i\sqrt{\frac{g}{N}}\lambda_{k}},\; k=1...N.
\end{equation}
In spite of the fact that we're integrating over real space
$\mathbb{R}^{N}$, the values $\lambda_k$ in the stationary point in
our case are complex
--- it is a usual situation for the steepest descent method. Let's now pass to the continuous case ($N\to\infty$), changing
$\lambda_{k}\to \sqrt{N} \lambda(x)$ ($x\in[0,1]$ is an analog of
the index $k$ for the continuous case). Then we get a system
\begin{eqnarray}
  {\rm P} \int_{C} \frac{v(\nu) d\nu}{\lambda-\nu} &=& \lambda + \frac{\sqrt{g}}{2} \frac{1}{\sqrt{g} \lambda +\frac{i}{2}} \\
  \int_{C} v(\nu) d\nu &=& 1,
\end{eqnarray}
where $v(\lambda)=\frac{dx}{d\lambda}$. From the discrete form
(\ref{m2}) of our equation one can see that, if
$\{\lambda_{k}\}_{k=1...N}$ is a solution, so is
$\{-\lambda_{k}^{\ast}\}_{k=1...N}$, therefore the contour $C$ is
symmetric with respect to the axis $Oy={\rm Im} \lambda$. Next we
multiply the first equation by $1-2i\sqrt{g}\lambda$ and introduce a
new function $\overline{v}(\lambda)=(1-2i\sqrt{g}\lambda)
v(\lambda)$. According to the position of the contour, $\int_{C}
\overline{v}(\lambda)\, d\lambda = D$ is real (because $\tr \lambda
= \int_{C} \lambda u(\lambda) d\lambda$ is imaginary). Then we get
an equation
\begin{equation}
{\rm P} \int_{C} \frac{\overline{v}(\nu) d\nu}{\lambda-\nu} =
i\sqrt{g} + \lambda (1-2i\sqrt{g}\lambda)
\end{equation}
We will look for a "resolvent" $\Phi(\lambda) \equiv
\int\limits_{C}\,\frac{\overline{v}(\nu) \,d\nu}{\lambda-\nu},\;
\lambda \notin C$ in the following form
\begin{equation}
\Phi(\lambda) = i\sqrt{g} + \lambda (1-2i\sqrt{g}\lambda) +
(A\lambda+B)\sqrt{(\lambda+b^{\ast})(\lambda-b)}
\end{equation}
In the limit $\lambda \to \infty$ the asymptotics is known
$\Phi(\lambda) \to \frac{D}{\lambda}$ (which is clear from the
definition of the resolvent and the normalization condition
--- see above), but since $D$ is unknown we will use the following conditions:
the coefficients of $\lambda^{2}, \lambda^{1}, \lambda^{0}$ vanish
and $\int_{C} v(\nu) d\nu = \int_{C} \frac{\overline{v}(\nu)
d\nu}{1-2i\sqrt{g}\nu}= 1$. The first three conditions can be
rewritten in the following form:
\begin{eqnarray}
  \lambda^{2}:&& -2i\sqrt{g} + A= 0 \\
  \lambda^{1}:&& 1+B+i\sqrt{g}(b^{\ast}-b)= 0 \\
  \lambda^{0}:&& \sqrt{g} ({\rm Re} b)^{2}+ B \,{\rm Im} b - \sqrt{g}=  0
\end{eqnarray}
In writing out these equations we took the positive value of the
square root on the continuation of the segment $[-b^{\ast};b]$ to
the right of $b$. Then on the axis, which is perpendicular to this
segment and intersects it in the middle, the square root is
imaginary with a positive (above the intersection point) or negative
(below the intersection point) imaginary part. Suppose $\lambda_{0}
\in C$. Taking into account what has been said and the
Sokhotsky-Plemelj formulas,
\begin{equation}
\underset{\epsilon\to 0}{{\rm
lim}}\;(\Phi(\lambda_{0}+i\epsilon)-\Phi(\lambda_{0}-i\epsilon))=2i(A\lambda_{0}+B)
\sqrt{(\lambda_{0}+b^{\ast})(b-\lambda_{0})}=-2i\pi\overline{v}(\lambda_{0}),
\end{equation}
thus
\begin{equation}
\overline{v}(\lambda) = -\frac{1}{\pi}(A\lambda+B)
\sqrt{(\lambda+b^{\ast})(b-\lambda)},\;\lambda \in C.
\end{equation}
Therefore we can rewrite the last condition in the form
\begin{equation}
-\frac{1}{\pi}
\int\limits_{-b^{\ast}}^{b}\,\frac{A\nu+B}{1-2i\sqrt{g}\nu}\,\sqrt{(\nu+b^{\ast})(b-\nu)}\,d\nu
= 1.
\end{equation}
So we're interested in the integrals of the following type:
\begin{eqnarray}
  I_{1} &=& \int\limits_{-b^{\ast}}^{b}\,\sqrt{(\nu+b^{\ast})(b-\nu)}\,d\nu \\
  I_{2} &=&
  \int\limits_{-b^{\ast}}^{b}\,\frac{d\nu}{1-2i\sqrt{g}\nu}\,\sqrt{(\nu+b^{\ast})(b-\nu)}.
\end{eqnarray}
They can be calculated in a similar fashion. Actually these
integrals are contour integrals along the upper bank of the cut from
left to right. Since on the lower bank of the cut the integrand
takes on an opposite value, the value of our integral is half the
value of the integral along a contour closed clockwise. The integral
along a closed contour in the first case $I_{1}$ is just $2\pi i
\times$(residue at infinity), and in the second case $I_{2}$ one
also has to take into account the residue at the pole. Adopting the
notation $\tau(\nu)=\sqrt{(\nu+b^{\ast})(b-\nu)}$,
\begin{eqnarray}
  I_{1} &=& \pi\,{\rm res}(\tau(\nu),\infty) \\
  I_{2} &=& \pi\,({\rm res}(\frac{\tau(\nu)}{1-2i\sqrt{g}\nu},\infty)+ {\rm
  res}(\frac{\tau(\nu)}{1-2i\sqrt{g}\nu},\nu_{0})),
\end{eqnarray}
were $\nu_{0}=\frac{1}{2i\sqrt{g}}$ is the position of the pole.
Upon calculating we obtain:
\begin{equation}
  I_{1} = \frac{\pi}{2} ({\rm Re} b)^{2}
\end{equation}
\begin{equation}
  I_{2} = \frac{\pi}{2\sqrt{g}} \left[\left(({\rm Re} b)^{2}+({\rm Im} b + \frac{1}{2\sqrt{g}})^{2}\right)^{1/2}-{\rm Im} b -
  \frac{1}{2\sqrt{g}}\right].
\end{equation}
Note that from the definition of the integral $I_{2}(g)$ it is clear
that it is nonsingular as $g \to 0$. Formally there are
singularities in the explicit expression obtained for $I_{2}$, but
it is easy to see that they cancel each other. So it is a check of
our calculations. Let's now write out all the conditions on
$(B,\,{\rm Re} \,b, \,{\rm Im} \,b)$, \mbox{eliminating $A$:}
\begin{eqnarray}
  B &=& -1 - 2 \sqrt{g} \, {\rm Im} b\\
  0 &=& \sqrt{g} ({\rm Re} b)^{2}+ B \,{\rm Im} b - \sqrt{g} \\
  1 &=& \frac{1}{\pi} I_{1} + \frac{2\sqrt{g} \, {\rm Im}
  b}{\pi}\,I_{2}
\end{eqnarray}
The equations are real, hereupon the number of unknowns coincides
with the number of equations. This confirms that it was quite
relevant to take into account the symmetry of the contour. Next we
introduce the parameter $\varepsilon = \frac{1}{2\sqrt{g}}$ and
eliminate $B$:
\begin{eqnarray}
  \frac{1}{2} &=& \frac{1}{2} ({\rm Re} b)^{2} - ({\rm Im} b)^{2} - \varepsilon \, {\rm Im} b   \\
  1 &=& \frac{1}{2} ({\rm Re} b)^{2} - ({\rm Im} b)^{2} - \varepsilon \, {\rm Im}b
  + {\rm Im} b\, \sqrt{({\rm Re} b)^{2} + ({\rm Im}b +
  \varepsilon)^{2}}.
\end{eqnarray}
Passing to the limit $\varepsilon \to 0$, introducing
$\overline{b}=\underset{\varepsilon \to 0}{{\rm lim}}\,
b(\varepsilon)$ and eliminating $({\rm Re} b)^{2}$ with the help of
the first equation, we get
\begin{equation}\label{s2.30}
\frac{1}{2} = {\rm Im} \overline{b}\,\sqrt{3({\rm Im}
\overline{b})^{2}+1}.
\end{equation}
The solution is as follows:
\begin{equation}
{\rm Im} \overline{b} = \frac{1}{\sqrt{6}};\;\;\; {\rm Re}
\overline{b} = \frac{2}{\sqrt{3}}.
\end{equation}
Thus, the contour $C$ lies above the real axis, so the pole
$\nu_{0}=-i \varepsilon$ never hits the contour, as one could
expect.
%\begin{figure}[h!]
%\begin{center}
%\includegraphics[height=7cm,width=7cm]{pic1.eps}
%\caption{} \label{picture}
%\end{center}
%\end{figure}
The ends of the contour in the limit $\epsilon \to 0$ turned out to
be on a finite distance from each other as well as from the origin.
It means that this approach correctly reflects the asymptotic
properties of the model
--- "collapsing" of the contour, which took place in (\ref{3.3})
as $g \to \infty$, is absent. Let's now return to the problem of
calculating the partition function. As it is clear from (\ref{3.5}),
the free energy in the limit $N\to\infty$ is:
\begin{equation}\label{m4}
\mathcal{E}=\int\,d\lambda\,v(\lambda)\,\left(\lambda^{2}+\ln(\frac{1}{2}-i\sqrt{g}\lambda)\right)-\int\,d\lambda\,d\mu\,v(\lambda)v(\mu)\,\ln{|\lambda-\mu|}.
\end{equation}
From the presented results one can see that there's a finite limit
of the function $v(\lambda)$, as $g\to\infty$. Therefore, the main
contribution to $\mathcal{E}$ in the described limit is given by
$\sqrt{g}$, which appears under the first logarithm, thus
\begin{equation}\label{m5}
\mathcal{E} \underset{g\to\infty}{\to}\,\frac{1}{2}\ln{g},
\end{equation}
which exactly coincides with the analogous expression for the vector
model (\ref{33}).
\section{Discussion and conclusion}
In this paper we have presented a detailed consideration of the
$D=0$ complex model. In particular, we obtained an expression for
the partition function by a method, analogous to the method
presented in \cite{brezin}. To the leading order in $g$ in the
strong coupling limit the hypothesis about the analogy of the matrix
and vector models turns out to be true. Moreover, it is also true in
the next-to-leading order if one uses in the vector model a finitely
renormalized charge $\widetilde{g}= \alpha \cdot g$ ($\alpha={\rm
const.}$) instead of $g$. This is clear from the general structure
of the free energy for both models:
\begin{equation}\label{nn}
E=\frac{1}{2}\,\ln{g}+\beta_{0}+\frac{\beta_{1}}{\sqrt{g}}+...
\end{equation}
We can always set the constant  $\beta_{0}$ to any value by changing
in an appropriate way $g\to \alpha\cdot g$. What is nontrivial is
the fact that the functional structures in the strong coupling
expansion of the matrix and vector models coincide, although the
coefficients may differ (a few first coefficients can be made equal
by an appropriate finite renormalization of parameters).

Our hypothesis is also true at $D=1$, which can be seen from the
explicit form of the free energy asymptotics for the matrix and
vector models: both behave like $E \sim \kappa\cdot g^{1/3}$
($\kappa = {\rm const}$). The question of whether our hypothesis is
true for $D\geq 2$ remains open. \acknowledgments{Our work was
partially financed by the Grant for the Support of Leading
Scientific Schools NSch-672.2006.1, the RFBR grant № 05-01-00541.
D.B. is grateful to Prof. M.M\"{u}ller-Preu{\ss}ker for hospitality
at the Humboldt University of Berlin, where a part of this work was
done, to the German Society for Academic Exchanges (DAAD) for
financial support in the framework of a Leonard Euler scholarship
and to the "Dynasty" foundation for financial support.}

\end{document}